\documentclass[sigconf]{acmart}

\usepackage{textcomp}
\usepackage{booktabs} 
\usepackage{algorithm}
\usepackage{algpseudocode}
\usepackage{enumitem}
\usepackage{amsmath}
\usepackage{subfig}
\usepackage{amssymb}
\usepackage{graphicx}
\usepackage{booktabs} 
\usepackage{url}  

\usepackage{enumitem}
\usepackage{graphicx}
\usepackage{multicol}
\usepackage[labelfont=bf]{caption}
\usepackage{subfig}
\usepackage{float}

\usepackage{multirow}
\usepackage{hhline}
\usepackage{booktabs} 
\usepackage{amssymb}
\usepackage{amsmath}

\usepackage{amsthm} 

\newtheorem{problem}{Problem}
\usepackage{algorithm}
\usepackage{algpseudocode}

\DeclareMathOperator*{\argmaxG}{argmax}   


\begin{document}
\title[Untraceable Web Browsing History and Unambiguous User Profiles]{Protecting User Privacy: An Approach for Untraceable Web Browsing History and Unambiguous User Profiles}

\author{Ghazaleh Beigi, Ruocheng Guo, Alexander Nou, Yanchao Zhang, Huan Liu}
\affiliation{%
	\institution{Computer Science and Engineering, Arizona State University, Tempe, Arizona}
}
\email{{gbeigi, rguo12, asnou, yczhang, huan.liu}@asu.edu}

%

\renewcommand{\shortauthors}{G. Beigi et al.}

\copyrightyear{2019} 
\acmYear{2019} 
\setcopyright{acmcopyright}
\acmConference[WSDM '19]{The Twelfth ACM International Conference on Web Search and Data Mining}{February 11--15, 2019}{Melbourne, VIC, Australia}
\acmBooktitle{The Twelfth ACM International Conference on Web Search and Data Mining (WSDM '19), February 11--15, 2019, Melbourne, VIC, Australia}
\acmPrice{15.00}
\acmDOI{10.1145/3289600.3291026}
\acmISBN{978-1-4503-5940-5/19/02}

\begin{abstract}
The overturning of the Internet Privacy Rules by the Federal Communications Commissions (FCC) in late March 2017 allows Internet Service Providers (ISPs) to collect, share and sell their customers' Web browsing data without their consent. With third-party trackers embedded on Web pages, this new rule has put user privacy under more risk. The need arises for users on their own to protect their Web browsing history from any potential adversaries. Although some available solutions such as Tor, VPN, and HTTPS can help users conceal their online activities, their use can also significantly hamper personalized online services, i.e., degraded utility.  In this paper, we design an effective Web browsing history anonymization scheme, \textsc{PBooster}, aiming to protect users' privacy while retaining the utility of their Web browsing history. The proposed model pollutes users' Web browsing history by automatically inferring \textit{how many} and \textit{what} links should be added to the history while addressing the utility-privacy trade-off challenge. We conduct experiments to validate the quality of the manipulated Web browsing history and examine the robustness of the proposed approach for user privacy protection.
\end{abstract}

%
%




\maketitle

\section{Introduction}
The web browsing history is the list of web pages a user has visited in past browsing sessions and includes the name of the web pages as well as their corresponding URLs. Online users usually expect a secure environment when surfing the Web wherein their personally identifiable information (a.k.a. PII) could be kept hidden from prying eyes. 
However, the web browsing history log is stored by the web browser on the device's local hard drive. In addition to the web browser, users' browsing histories are recorded via third-party trackers embedded on the web pages to help improve online advertising and web surfing experience. 
Moreover, Internet Service Providers (ISPs) such as AT\&T and Verizon, have full access to individuals' web browsing histories. ISPs can infer different types of personal information such as users' political views, sexual orientations and financial information based on the sites they visit. Some countries have policies for protecting individuals' privacy. For example, European Union (EU) has regulated a new data protection and privacy policy for all individuals within the European Union and the European Economic Area (a.k.a. General Data Protection Regulation (GDPR)).\footnote{https://bit.ly/1lmrNJz}United States government also had Federal Communications Commission's (FCC) landmark Internet privacy protections for users such that ISPs could have been punished by the Federal Trade Commission (FTC) for violating their customers' privacy. However, not all countries have such policies. FCC's Internet privacy protection has been also removed in late March of 2017. This new legislation allows ISPs to monitor, collect, share and sell their customer's behavior online such as detailed Web browsing histories without their consent and any anonymization.\footnote{http://wapo.st/2mvYKGa}

Assuming that ISPs and online trackers make browsing history data pseudonymous before sharing, a recent study has shown the fingerprintability of such data by introducing an attack which maps a given browsing history to a social media profile such as Twitter, Facebook, or Reddit accounts~\cite{su2017anonymizing}. 
Although linking browsing history to social media profiles may not always lead to figuring out one's real identity, it is a stepping stone for attackers to infer real identities. This identity exposure may result in harms ranging from persecution by governments to targeted frauds~\cite{christin2010dissecting,beigi2018securing}.

The onus is now on the users to protect their browsing history from any kind of adversaries like ISPs and online trackers. There are approaches to help users shield their web browsing history such as browser add-ons or extensions (e.g., `Ghostery', `Privacy Badger' and `HTTPS everywhere'), Virtual Private Networks (VPN) services, Tor, and HTTPS. 
However, none of the above solutions can prevent ISPs from collecting users' web browsing history and protect users' identities when such information is revealed because de-anonymization attacks will still work~\cite{su2017anonymizing}. Moreover, using these solutions could result in a severe decrease in the quality of online personalization services due to the lack of customer's information. This information is critical for online vendors to profile users' preferences from their online activities to predict their future needs. So users face a dilemma between user privacy and service satisfaction. Hereafter, we refer to a user's satisfaction of online personalization services, as \textit{online service utility}, or simply, \emph{utility}. The aforementioned challenges highlight the need to have a web browsing history anonymizer framework, which can help users strike a good balance between their privacy and utility. Traditional privacy preserving web search techniques such as~\cite{yang2016privacy,zhang2016anonymizing,zhu2010anonymizing} are designed for different purposes and are thus ineffective in accomplishing our goals. 

Intuitively, the more links we add to a web browsing history, the more privacy we can preserve. An extreme case is when the added links completely change a user's browsing history to perfectly obfuscate the user's fingerprints. Some existing methods include ISPPolluter,\footnote{https://github.com/essandess/isp-data-pollution} Noiszy,\footnote{https://noiszy.com/} and RuinMyHistory\footnote{https://github.com/FascinatedBox/RuinMyHistory} which pollute a web browsing history by adding links randomly. However, such methods largely disturb user profiles and thus results in the loss of utility of online services. Similarly, the maximum service utility can only be achieved at the complete sacrifice of user privacy.
It is challenging to design an effective browsing history anonymizer that retains high utility. 
In this paper, we aim to study the following problem: \textit{how many} links and \textit{what} links should be added to a user's browsing history to boost user privacy while retaining high utility. 
Note that links cannot be removed from the browsing history as all of user's activities have been already recorded by ISPs. The research requires quantifying the privacy of users and the utility of their services. We address these challenges within a novel framework, called \textsc{PBooster}. This framework exploits publicly available information in social media networks as an auxiliary source of information to help anonymizing web browsing history while preserving utility. Our contributions can be summarized as follows.
\begin{itemize}[leftmargin=*]
	\item We address the problem of anonymizing web browsing histories while retaining high service utility. We show that this problem cannot be solved in polynomial time.
	\item We propose an efficient framework, \textsc{PBooster}, with measures for quantifying the trade-off between user privacy and the quality of online services. 
	\item We conduct experiments and evaluate the proposed approach in terms of privacy and utility. 
	Results demonstrate the efficiency of \textsc{PBooster} in terms of privacy-utility trade-off. 
\end{itemize}

\section{Related Work}

Explosive growth of the Web has not only drastically changed the way people conduct activities and acquire information, but also has raised security~\cite{alvari2018early,alvari2017semi} and privacy~\cite{narayanan2009anonymizing,beigi2018securing,beigi2018privacy} issues for them. Identifying and mitigating user privacy issues has been studied from different aspects on the Web and social media (for a comprehensive survey see~\cite{beigi2018privacy}). Our work is related to a number of research which we discuss below while highlighting the differences between our work and them. 

\noindent\textbf{Tracking and Profiling.}
In the area of user tracking and profiling, there have been efforts to study how and to what degree that web browser tracking~\cite{lerner2016internet,englehardt2016online,meng2016trackmeornot} and cross-device tracking~\cite{zimmeck2017privacy} can be done by third parties.
This line of work mainly studied the mechanisms of the user tracking techniques.
To go one step further, for the protection of users from tracking and profiling, strategies such as limiting the access to sampled user profiles~\cite{singla2014stochastic} and distorting the user profile~\cite{castella2009preserving} are proposed to minimize privacy risk. 

\noindent\textbf{Privacy Preserving Web Search.}
Web search has become a regular activity where a user composes a query formed by one or more keywords and sends it to the search engine. The engine returns a list of web pages according to the user query. These search queries are a rich source of information for user profiling. Privacy preserving web search approaches focus on anonymizing users search queries. One group of works focused on the protection of post-hoc logs~\cite{cooper2008survey,gotz2012publishing,zhang2016anonymizing}. 
Another group of approaches including client-side ones focuses on search query obfuscation~\cite{ye2009noise,balsa2012ob,gervais2014quantifying,howe2009trackmenot}. These approaches are user-centric and automatically generate fake search queries on behalf of user. 
%
%
%
 Web browsing history anonymization problem is different from privacy preserving web search problem. The former consists of a set of URLs a user has visited in past browsing sessions, while the latter includes a set of queries and relevant pages returned by search engine for each given query. Moreover, in web browsing history anonymization, URLs cannot be removed from a user's history (all activities have been already recorded by ISPs) while a data publisher is allowed to remove a portion of queries and pages in privacy preserving web search problem. This makes the web browsing history anonymization more challenging.

\noindent\textbf{Privacy Preserving Recommendation.}
Recommendation systems help individuals find relevant information, however, these systems can also raise privacy concerns for users~\cite{beigi2018similar}. 
Differential privacy based approaches~\cite{mcsherry2009differentially,machanavajjhala2011personalized,shen2014privacy} add noise to recommendation results so that the distribution of results is insensitive to the records of any specific user. 
%
Secure computation approaches have been also designed to take care of the computation procedure in recommender systems.
Privacy-preserving matrix factorization schemes~\cite{hua2015differentially} are designed to avoid exposure of user information during the recommender's computation. Matrix factorization has been used in many applications such as recommendation systems and trust/distrust prediction~\cite{beigi2016exploiting}.
Another work~\cite{krause2010utility} studies sharing user attributes to recommenders while handling the trade-off between privacy and quality of received service. Recent work~\cite{shen2016epicrec} also studied utility-privacy trade-off. 

The problem of anonymizing web browsing history is unique in this work. 
First, in our problem web browsing URLs cannot be removed and the original format of data will be published rather than its statistics. This also makes differential privacy based techniques ineffective for this task. Second, the user is not aware of the tasks that use his data and thus securing computation approaches is impractical for this new problem. Third, the proposed solution should be efficient and results in minimal loss in utility. All of these make this problem even more challenging.


\section{Threat Model and Problem Statement}
Before discussing the details of the proposed solution, we first formally define browsing history, then review the web browsing history de-anonymization and finally introduce the problem of web browsing history anonymization. For each user, web browsing history is defined as the list of web pages a user has visited in his past browsing sessions and includes the corresponding URLs of the visited web pages. This log is recorded by the browser, third-party trackers and ISPs. In addition to his browsing history, other private data components such as cache, cookies and saved passwords are also saved during a browsing session which are sometimes referred to under the browsing history umbrella. However, in this work, we separate these pieces of information from web browsing history. Given a user $u$, we assume his web browsing history $\mathcal{H}^u$ is generated by a sequence of $n$ links $\mathcal{H}^u=\{l_1,...l_n\}$ where $l_i$ corresponds to the $i$-th URL visited by the user $u$. 
\subsection{Threat Model}\label{section:threat}
 De-anonymizing browsing histories is a type of linkage attack which is introduced by Su et al.~\cite{su2017anonymizing}. This de-anonymization attack links web browsing histories to social media profiles. The main idea behind this threat model is that people tend to click on the links in their social media feed. These links are mainly provided by the set of user's friends. Since each user has a distinctive set of friends on social media and he is more likely to click on a link posted by any of his friends rather than a random user, these distinctive web browsing patterns remain in his browsing history. Assuming that the attacker knows which links in the history have resulted from clicks on social media feeds, a maximum likelihood based framework is developed as a de-anonymization attack which identifies the feed in the system that has more probably generated the browsing history. This attack can be formally defined as:
\begin{problem}
	Given user $u$'s web browsing history $\mathcal{H}^u = \{l_1, ...l_n\}$ which is consisted of $n$ links, map $u$ to a social media profile whose feed has most probably generated the browsing history~\cite{su2017anonymizing}.
\end{problem}

Let's assume that each user $u$ has a personalized set of recommender links. For example, this recommendation set could be a set of links appeared in the user's social media Feed (e.g., Twitter) which includes links posted by the user's friends on the network. Su et. al.~\cite{su2017anonymizing} assume that each user visits links in his recommendation set. Given a browsing history $\mathcal{H}^u$, the attacker finds the most likely recommendation set that corresponds to the given user $u$: the recommendation set which contains many of the URLs in the browsing history and is not too big. This de-identifies the browsing history. For the detailed proof and implementation of this attack please refer to~\cite{su2017anonymizing}. Twitter is selected as a mapping platform for evaluation of this attack. This work shows that users' activities in social media can be used to re-identify them. 
We next introduce the problem of web browsing history anonymization.

\subsection{Problem Statement}
In this work, we define a privacy preserving framework which protects user's privacy by combating the de-anonymizing web browsing history threat model we discussed in Section \ref{section:threat}. In addition, utility here is also defined as user's satisfaction of online personalized services. This could also be measured by comparing the quality of manipulated web browsing history after anonymization with the original one. Given user $u$'s browsing history $\mathcal{H}^u$, the goal is to anonymize $u$'s browsing history by adding new links to $\mathcal{H}^u$ in an efficient manner, such that both the user's privacy and utility are preserved, i.e., web browsing history is robust against de-identification attack and maintains its utility. 

We first need to convert links to a structured dataset. One straightforward solution is to leverage the content of each web page and then map it to a category or a topic selected from a predefined set. This way, each user will be represented by a set of categories extracted from all of the web pages he has visited. 
One typical way for extracting topics is to manually define them (e.g., sports, fashion, knowledge, etc.) and then map each web page to the corresponding category. This method requires a set of keywords related to each topic and then inferring the web page's topic by calculating the similarity of its textual content to the given keywords. This solution is not feasible in practice since it needs frequent updates of keywords for each category due to the fast growth of the Internet. Moreover, this only provides a coarse-grained categorization of web pages' contents. In order to have a finer level of granularity we follow the same approach as in~\cite{phuong2014gender} and adopt Latent Dirichlet Allocation (LDA) topic modeling technique ~\cite{blei2003latent}. We use the following procedure to assign topics for each web page:
\begin{enumerate}[leftmargin=*]
	\item We retrieve a set of web pages to construct a corpus and then use LDA to learn topic structures from the corpus. 
	\item For each web page, the learned topic model in the previous step is used to infer the topic proportion and topic assignment based on the textual content of the page.
	\item The topic with highest probability from the topic distribution is selected as the representative topic of the page.
\end{enumerate}

We use $\mathcal{T} = \{t_1,..., t_m\}$ to denote the set of learned topics. Then each link in the browsing history $\mathcal{H}^u$ is mapped to a topic in the topic set, $t_l \in \mathcal{T}$. Matrix $\mathbf{T}^u \in \mathbb{R}^{n\times m}$ is then used to represent the link-topic relationship for all the links in $\mathcal{H}^u$ where $\mathbf{T}^u_{ij} = 1$ indicates that $i$-th link of user $u$ is correlated to the topic $t_j$. The problem of anonymizing browsing history of user $u$ is then formally defined as:
\begin{problem}
	Given user $u$'s browsing history $\mathcal{H}^u$, and link-topic matrix ${\mathbf{T}^u}$, we seek to learn an anonymizer $\mathit{f}$ to create a manipulated browsing history $\widetilde{\mathcal{H}^u}$ by adding links to $H^u$ to preserve the privacy of user $u$ while keeping the utility of $\widetilde{\mathcal{H}^u}$ for future applications.
	\begin{align}
		f:\{ \mathcal{H}^u,{\mathbf{T}^u}\} \rightarrow \{\widetilde{{\mathcal{H}^u}}\}
	\end{align}
\end{problem}
We stress that links cannot be removed from the browsing history as all of user's activities have been already recorded by ISPs.

\section{A Framework for Privacy Boosting}
The goal of the web browsing history anonymizer is to manipulate the user's browsing history by adding links in a way that: 1) user privacy is preserved even when the adversary publishes the data with the weakest level of anonymization (i.e., just removing PIIs) and 2) browsing history still demonstrates user's preferences so that the quality of personalized online services is preserved.

An immediate solution that may come to mind is to add links from popular web sites. This approach cannot preserve privacy as the adversary can easily remove popular links from the history and then deploy the attack. Another solution could be adding links from the browsing history of users who are very similar to $u$, i.e., his friends in social media. This approach can preserve the utility of browsing history but fail to make the user robust to the adversary attack. This is also observed in~\cite{su2017anonymizing} where it was shown that the more a user's history contains links from his friends' browsing activities in social media, the more fingerprints he leaves behind. 
All these emphasize the need for an effective solution which can handle the utility-privacy trade-off. 

In this section we will discuss how our proposed algorithm \textsc{PBooster}, can handle utility-privacy trade-off. To better guide the \textsc{PBooster} and to assess the quality of the altered history, we need measures for quantifying the effect of adding links on user privacy and utility. We first present these measures and then detail the \textsc{PBooster}.

\subsection{Measuring User Privacy}
The best case for user privacy is when a user's visited links (i.e., interests) are distributed uniformly over different topics. This improves the user privacy by increasing ambiguity of his interests distribution. This makes it harder for the adversary to infer the real characteristic of the user's preferences and then re-identify him by mapping his anonymized information to a real profile. Entropy is a metric which measures the degree of ambiguity. We leverage the entropy of the user's browsing history distribution over a set of predefined topics as a measure of privacy. 

We first introduce the topic-frequency vector $\mathbf{c}_u \in \mathbb{R}^{m \times 1}$ as $\langle c_{u1}, c_{u2}, ..., c_{um}\rangle$ for each user $u$, where $c_{uj}$ is the number of links in $u$'s history related to the topic $t_j$. Note that $\sum_{j=1}^{m} c_{uj} = \lvert \mathcal{H}^u\rvert$ where $\lvert.\rvert$ denotes the size of a set. The topic probability distribution for each user can be then defined as $\mathbf{p}_u = J(\mathbf{c}_u) = \langle p_{u1}, p_{u2}, ..., p_{um}\rangle$ where $J$ is the normalization function of input vector $\mathbf{c}_u$ where $p_{uj} = \frac{c_{uj}}{\lvert{H}^u\rvert}$ and $\sum_{j=1}^{m} p_{uj} = 1$. The privacy of user $u$, which is the degree of ambiguity of his browsing history, can be captured by the entropy of the topic probability distribution $\mathbf{p}_u$. This measures the spread of the user's interests across different topics. Given topic probability distribution, privacy is measured as:
\begin{equation}\label{eq:privacy}
Privacy({p_u})= -\sum_{j = 1}^{m} {p_{uj}}\log {p_{uj}}
\end{equation}
The higher this metric is, the greater the user privacy. The optimal value of this measure is thus achieved when the user's browsing links topics are distributed uniformly across the set of topics. 

\subsection{Measuring Utility Loss}
Utility or quality of online services is a measurement of a user's satisfaction from the online personalized services he receives based on his online activities. This measurement should be able to estimate the loss of quality of services after manipulating the user's browsing history by the \textsc{PBooster}. We quantify utility loss as the difference between a user's topic distribution before and after browsing history manipulation. Finding the difference between topic distributions has been exploited in other applications such as recommender systems~\cite{li2011scene}. We use the same notion used in~\cite{li2011scene} 
and quantify the utility loss between $\mathbf{p}_u$ and $\hat{\mathbf{p}_u}$ as:
\begin{equation}\label{eq:uitlity}
utility\textunderscore {loss}(\mathbf{p}_u,\hat{\mathbf{p}_u}) =0.5\times (1-sim(\mathbf{p}_u,\hat{\mathbf{p}}_u))
\end{equation}
where $\hat{\mathbf{p}_u}$ denotes the new topic probability after manipulating history. One typical choice for the $sim$ is cosine similarity~\cite{li2011scene}: 
\begin{align}
	sim(\mathbf{p}_u,\hat{\mathbf{p}_u})  = \frac{\mathbf{p}_u . \hat{\mathbf{p}}_u}{\lVert \mathbf{p}_u\rVert.\lVert\hat{\mathbf{p}}_u\rVert}
\end{align}

Since $sim \in [-1,1]$, the output of $utility\textunderscore {loss}$ function will be in $[0,1]$. According to this measure, the minimum value for utility loss is when $\mathbf{p}_u = \hat{\mathbf{p}}_u$ and the maximum is reached when $\hat{\mathbf{p}}_u$ does not have any non-zero value in common with $\mathbf{p}_u$. 

\subsection{\textsc{PBooster} Algorithm}
We have discussed so far how to quantify a user's utility and privacy according to his browsing history. The goal is now to find a set of new links $\mathcal{A}$ to add to the browsing history such that, 1) $privacy(\hat{\mathbf{p}_u})$ is as large as possible, and 2) $utility\textunderscore {loss}(\mathbf{p}_u,\hat{\mathbf{p}_u})$ is as small as possible. However, as we discussed earlier, the optimal value for privacy is reached when the user's interests are spread uniformly across different topics, while the utility loss is minimized when no changes have been done to the topic distribution ${\mathbf{p}_u}$. This raises a trade-off issue between user's privacy and utility loss. Simply put, maximizing privacy results in the loss of utility and vice versa. In order to optimize the trade-off between utility loss and privacy for each user $u$, we define a new scalar objective function:
\begin{align}
	G(J(\mathbf{c}_u),J(\hat{\mathbf{c}_u}),\lambda) = \lambda* privacy(J(\hat{\mathbf{c}_u}))
	- utility\textunderscore {loss}(J(\mathbf{c}_u),J(\hat{\mathbf{c}_u}))\label{G}
\end{align}
where $\hat{\mathbf{c}_u}$ is the topic-frequency vector after manipulating browsing history and $\lambda$ controls the contribution of privacy in $G$. We aim to find a set of links $\mathcal{A}$ by solving the following optimization problem:
\begin{align}\label{optimization}
	\mathcal{A^*} = \argmaxG_{\mathcal{A}} G(J(\mathbf{c}_u),J(\hat{\mathbf{c}_u}),\lambda)
\end{align}
where $\hat{\mathbf{c}_u}$ could be made from $\widetilde{{\mathcal{H}^u}}= \mathcal{H}^u\cup \mathcal{A}$. Topic distribution $\hat{\mathbf{p}_u}$ is constructed from $\hat{\mathbf{c}_u}$ accordingly. It's notable to say that the value of $\lambda$ has impact on the inferred set of links $\mathcal{A^*}$ in a sense that larger values of $\lambda$ will lead to a browsing history $\widetilde{{\mathcal{H}^u}}$ with higher privacy while lower $\lambda$ values result in lower utility loss. 

It is worthwhile to mention that the search space for this problem (Eq.\ref{optimization}) is exponential to $N$ ($\mathcal{O}(m\times 2^N)$), where $N$ is the maximum of the number of links w.r.t. a topic. Considering this fact, it can be expensive and even infeasible to search for the optimal solution. We thus decide to approach this problem in an alternative way. We divide the optimization problem in Eq.\ref{optimization} into two subproblems :
\begin{enumerate}[leftmargin=*]
	\item \textbf{Topic Selection: }Selecting a subset of topics and calculating the number of links which should be added to each topic in order to maximize the function $G$ as follows:
	\begin{align}\label{optimization2}
		{a^*} = \argmaxG_{{a}} G(J(\mathbf{c}_u),J(\hat{\mathbf{c}_u}),\lambda)
	\end{align}
	where $\mathbf{a} = \langle a_{1},..,a_{m}\rangle  \in \mathbb{R}^{m \times 1}$ such that each non-zero element $a_{i}$ indicates the \textit{number of to-be added new links} which are related to the topic $t_i$. Zero value means that none of the new links are associated with the topic $t_i$. Consequently, $\hat{\mathbf{c}_u}$ is defined as $\hat{\mathbf{c}_u} = \langle c_{u1}+ a_{1},..,c_{um}+a_{m}\rangle $. 
	This step indicates the number of links which should be added to each topic to maximize $G$.
	\item \textbf{Link Selection: }Selecting a proper set of links which corresponds to the identified topics and their numbers found in the previous step.
\end{enumerate}
To recap, the \textsc{PBooster} algorithm anonymizes a user's browsing history by first selecting a subset of topics with the proper number of links for each topic (topic selection phase) and then finding corresponding links for each of them (link selection phase). Next, we will discuss the possible solutions for each step.
\subsection{Topic Selection} 
One brute-force solution to the optimization problem in Eq.\ref{optimization2}, is to evaluate all possible combinations of a set of topics with different sizes to find the best ${\mathbf{a}^*}$. The exponential computational complexity of this algorithm makes it unacceptable and even impractical when quick results are required. We thus need a more efficient solution.

According to a recent study~\cite{Guerraoui:2017:UPE:3077136.3080783}, having more information in the browsing history will not necessarily increase either the utility or the privacy. In other words, with large information available on user's preferences, observing a new link would have little to no impact on enhancing utility and privacy of the user. Simply put, adding more data to the history, could make the user less secured, with no specific improvement observed in the utility. The submodularity concept formally captures this intuition. A real valued function $f$ is submodular if for a finite set $\mathcal{E}$ and two of its subsets $\mathcal{X}$, $\mathcal{Y}$ where $\mathcal{X}\subseteq \mathcal{Y} \subseteq \mathcal{E}$, and $e \in \mathcal{E}\backslash Y$, the following property holds:
\begin{align}
	f(\mathcal{X}\cup \{e\}) - f(\mathcal{X})\geq f(\mathcal{Y}\cup \{e\})-f(\mathcal{Y})
\end{align}

This means that adding one element $\{e\}$ to the set $\mathcal{X}$ increases $f$ more than adding $\{e\}$ to the set $\mathcal{Y}$ which is superset of $\mathcal{X}$~\cite{nemhauser1978analysis}. This intuitive diminishing return property exists in different areas such as social media networks and recommender systems. Recall from Eq.~\ref{G} that the function $G$ is consisted of two components, namely privacy and utility loss. Given $\lambda \in [0,1]$ and topic-frequency vector $\mathbf{c}_u$, we can rewrite the optimization problem in Eq.\ref{optimization2} as:
\begin{equation}
\label{eq:NIP}
\begin{split}
\underset{\mathbf{a}}{\argmaxG}  -\lambda(\sum_{j}\hat{p}_{uj}\log&\hat{p}_{uj}) -  0.5\times(1-\frac{\sum_j p_{uj}\hat{p}_{uj}}{\sqrt{\sum_j p_{uj}^2}\sqrt{\sum_j\hat{p}_{uj}^2}}) \\
subject \; to \;  -\hat{c}_{uj} & \le -c_{uj} \;,  \hat{c}_{uj} \in \mathbb{N}_0\\
\end{split}
\end{equation}  
where $\hat{p}_{uj} = \frac{\hat{c}_{uj}}{|\widetilde{{\mathcal{H}^u}}|}$ is the topic probability distribution after applying \textsc{PBooster}. Privacy is calculated using the entropy function which is submodular in the set of random variables~\cite{krause2008near}. The defined utility loss is also naturally submodular~\cite{li2011scene}. Since nonnegative linear combinations of submodular functions are submodular as well, the objective function $G$ is submodular. $G$ is also non-monotone and thus the problem in Eq.\ref{eq:NIP} is equal to maximizing a non-monotone nonnegative submodular function. This problem has been shown to be NP-hard~\cite{feige2011maximizing} and there is no optimal solution for it in an efficient amount of time.



However, the problem of maximizing non-monotone non-negative submodular function has been solved earlier~\cite{feige2011maximizing}. A greedy local search algorithm, LS, has been introduced for solving this problem which was proved to guarantee a near-optimal solution. The greedy LS achieved a value of at least $\frac{1}{3}$ of the optimal solution~\cite{feige2011maximizing}. Formally speaking, if we assume solution $\mathbf{a}_G$ is provided by the greedy LS algorithm, and $\hat{\mathbf{c}_G} = \mathbf{c}_u + \mathbf{a}_G$, and the optimal solution is $a_{\textsc{OPT}}$, and $\textsc{OPT}(\mathbf{c}_u) = \mathbf{c}_u + \mathbf{a}_{\textsc{OPT}}$, the following theorem holds:
\begin{theorem}\label{theo2} {\it If $G(.,.)$ is a nonnegative non-monotone submodular function, the set of topics $\mathbf{a}_G$ found by the greedy algorithm has the following lower bound~\cite{feige2011maximizing}:}
	\begin{align}
		G(J(\mathbf{c}_u),J(\hat{\mathbf{c}_u}),\lambda)
		\geq (\frac{1}{3}-\frac{\epsilon}{n}) 	G(J(\mathbf{c}_u),J(\textsc{OPT}(\mathbf{c}_u)),\lambda) 
	\end{align}
\end{theorem}
 Here, $\epsilon > 0$ is a small number. Local search algorithm iteratively adds an element to the final set or removes one from it to increase the value of $G$ until no further improvement can be achieved. Algorithm \ref{alg:alg1} shows the topic selection algorithm which deploys the greedy local search. Elements of $\mathbf{a} = \langle a_{1},.., a_{m}\rangle$ will be increased or decreased iteratively to increase value of $G$ until it cannot be improved anymore.

\begin{algorithm}[t]
	\caption{\textbf{Greedy local search for topic selection}}\label{alg:alg1}
	
	\begin{algorithmic}[1]
		\Require \text{ topic-frequency vector $\mathbf{c}_u$, $\lambda$, $\epsilon$}
		\Ensure $\mathbf{a} = \langle a_{1}, a_{2}, ..., a_{m}\rangle$
		\State \text{Initialize $\mathbf{a} = \langle 0, 0, ..., 0\rangle$},  $\hat{\mathbf{c}_u} = \mathbf{c}_u + \mathbf{a}$ and $val \longleftarrow 0$
		\While {\text{there is increase in in value of $G(J(\mathbf{c}_u),J(\hat{\mathbf{c}_u}),\lambda)$}}
		\State Select $t_j, j\in\{1,...,m\}$ such that by updating $a_{j} \longleftarrow a_{j}+1$ and $\hat{\mathbf{c}_u} = \mathbf{c}_u + \mathbf{a}$, then $G(J(\mathbf{c}_u),J(\hat{\mathbf{c}_u}),\lambda)$ is maximazed
		\State Update \text{$val \longleftarrow G(J(\mathbf{c}_u),J(\hat{\mathbf{c}_u}),\lambda)$}
		\If {$\exists~ t_j$ such that updating $a_{j} \longleftarrow a_{j}+1$ and $\hat{\mathbf{c}_u} = \mathbf{c}_u + a$ results in $G(J(\mathbf{c}_u),J(\hat{\mathbf{c}_u}),\lambda) > (1+\frac{\epsilon}{n^2}) .~val$}
		\State Update $a_{j} \longleftarrow a_{j}+1$ , \text{$val \longleftarrow G(J(\mathbf{c}_u),J(\hat{\mathbf{c}_u}),\lambda)$}
		\State Repeat from step 5
		\EndIf
		\If {$\exists~ t_j$ such that  $a_{j}\geq 1$ and updating $a_{j} \longleftarrow a_{j}-1$ and $\hat{\mathbf{c}_u} = \mathbf{c}_u + \mathbf{a}$ results in $G(J(\mathbf{c}_u),J(\hat{\mathbf{c}_u}),\lambda) > (1+\frac{\epsilon}{n^2}) .~val$}
		\State Update $a_{j} \longleftarrow a_{j}-1$ , \text{$val \longleftarrow G(J(\mathbf{c}_u),J(\hat{\mathbf{c}_u}),\lambda)$}
		\State Repeat from step 5
		\EndIf
		\EndWhile
	\end{algorithmic}
\end{algorithm}

We emphasize that according to~\cite{feige2011maximizing}, there is no efficient algorithm which could select the best set of links to maximize aggregation of both privacy and utility in polynomial time. Following the Theorem~\ref{theo2}, the proposed greedy algorithm can select a set with a lower bound of $\frac{1}{3}$ of the optimal solution, providing the maximum user privacy and utility in polynomial time.
\subsection{Link Selection}
Previously, we discussed the solution for selecting a subset of topics and the proper number of links for each topic to preserve user privacy while keeping the new topic distribution as close as possible to the original one. The second step in \textsc{PBooster} is to select links which correspond to the selected set of topics.
Let us assume that user $u$ has at least one active\footnote{Here, user activity does not refer to posting contents. In this work, we assume a user as active if he visits his feed and have non-empty list of friends.} account on a social media site and \textsc{PBooster} has access to the list of user's friends $\mathcal{F}^u \neq \emptyset$. 


We propose the following solution for the link selection problem. For each single update $\omega$ in the vector $a$, we randomly select a user $v$ with a public social media profile from outside of the list of $u$'s friends, $v \notin \mathcal{F}^u$. We then simulate $v$'s browsing history $\mathcal{H'}_v$, with the size of $|\mathcal{H'}_v|=q$. The detail of this simulation is discussed in the next section. Link-topic relation matrix $\mathbf{T}^v$ will be constructed from the history $\mathcal{H'}_v$. If there is no link in $\mathcal{H'}_v$ which corresponds to the topic of $\omega$, then the process will be repeated for another random user, otherwise, a random related link will be chosen. The pseudocode of this algorithm is shown in Algorithm \ref{alg:alg2}. 

\begin{algorithm}[t]
	\caption{\textbf{Link selection}}\label{alg:alg2}
	
	\begin{algorithmic}[1]
			\Require \text{$\mathcal{F}^u$, q, $\mathbf{a} = \langle a_{1}, a_{2}, ..., a_{m}\rangle$}
			\Ensure \text{Set of links $\mathcal{A}$}
			\State \text{$\mathcal{A} = \emptyset$}
			\For {\text{each update $\omega$ in $a$}}
			\State \text{Let $t_j$ be the corresponding topic of update $\omega$}
			\State \text{Select a user $v$ randomly such that $v \notin \mathcal{F}^u$}
			\State Simulate a browsing history $\mathcal{H'}_v$ for $v$ with the size of $q$. Make $c_v$ and link-topic matrix $\mathbf{T}^u$ from $\mathcal{H'}_v$
			\If {\text{$c_{vj} = 0$}}	 \text{: Go to line 4 and repeat, \textbf{else}}
			\State Select a non-zero row $r$ randomly from $\mathbf{T}^v[:,j]$
			\State \text{Select corresponling link $l$ to row $r$}
			\State \text{$\mathcal{A} = \mathcal{A} \cup \{l\}$}
			\EndIf
			\EndFor
	\end{algorithmic}
\end{algorithm}

To recap, \textsc{PBooster} uses the greedy local search algorithm for submodular maximization to first find the topics which need to be updated and then infer the number of links which should be added to those topics in a way that user privacy and utility is maximized. 
\section{Experimental Evaluation}
In this section we conduct experiments to evaluate the effectiveness of \textsc{PBooster} in terms of both privacy and utility. In particular, we seek to answer the following questions: (1) how successful is the proposed defense in protecting users' privacy? (2) how does \textsc{PBooster} affect the quality of online services? (3) how successful is \textsc{PBooster} in handling privacy-utility trade-off?
\subsection{Dataset}
Su et al.~\cite{su2017anonymizing} evaluate their de-anonymization strategy by examining it on a set of synthetically generated histories as well as real, user-contributed web browsing histories. 
Synthetic history is generated for a set of users based on their activities in social media. These users are selected semi-randomly from social media real-time streaming API-- the more active a user is, the more likely he is to be chosen. The histories are simulated in a way that mimic users' real online behaviors--they mostly click on links posted to their news feed, and sometimes click on links posted by their friends-of-friends~\cite{su2017anonymizing}. These friends-of-friends links may be clicked due to the organic exploration behavior of people or the Social media's algorithmic recommendation system that tries to get users visit their friends-of-friends links~\cite{su2016effect}. Their results on real user generated browsing history is consistent with the results of synthetic histories. This confirms the procedure of simulating synthetic browsing history as well as the efficiency of the generated data~\cite{su2017anonymizing}.

Similar to Su et al~\cite{su2017anonymizing}, we examine the performance of \textsc{PBooster} on a set of synthetically generated browsing history. We follow the same procedure as in~\cite{su2017anonymizing} to simulate the browsing history dataset. To generate the synthetic history for each user $u$, friend's links and friends-of-friends' links are generated accordingly~\cite{su2017anonymizing}. Friends' links are generated by pulling links from a randomly selected friend of $u$. Friends-of-friends' links are also generated by first picking one of $u$'s friends, say $v$, uniformly at random, and then pulling a link from one of $v$'s friends. Following~\cite{su2017anonymizing}, we select Twitter as the source of users' activities to simulate data because of two reasons. First, many users activities on Twitter are public, and second Twitter has real-time API which helps avoid the need for large-scale web crawling. We select a total number of $1200$ users and following~\cite{su2017anonymizing}, we generate histories of various sizes including $\{30, 50, 100\}$ for each user. For each history, $16\%$ of links are from friends-of-friends and the rest are from friends. Note that we only select links that are related to web pages in English to make the textual analysis easier.
\subsection{Experiment Setting}
To simulate the real-world browsing situation, we divide the browsing history into $\frac{|\mathcal{H}^u|}{h}$ batches of links with size of $h$. These batches will be added to the history incrementally and \textsc{PBooster} will anonymize the updated history after taking each batch. We set the values $h=25$, $q=20$ (used in link selection algorithm) and trade-off coefficient $\lambda = \{0,0.1,0.5,1, 10,20,50,70,$ $100\}$. We use LDA topic modeling from Python package gensim~\cite{rehurek2010software} and set the number of topics $m=20$ and LDA parameters $\alpha=0.05, \beta=0.05$. We compare \textsc{PBooster} with the following baselines:
\begin{itemize}[leftmargin=*]
	\item \textsc{Random}: Assuming $x$ new links are added by \textsc{PBooster}, this approach selects $x$ links randomly from the browsing history of users who are not from $u$'s friends. Note that this method does not consider the topics of the links. We compare our model against this method to investigate whether the topics of the chosen links will have effect on the privacy of the users, or in other words, how well topic selection technique in Algorithm~\ref{alg:alg1} performs?
	\item \textsc{JustFriends}: This approach is quite similar to \textsc{PBooster} except that in the link selection phase, it adds links from a user's friends' simulated browsing history. We use this method to see how well our link selection technique in Algorithm~\ref{alg:alg2} performs.
	\item \textsc{ISPPolluter\footnote{https://github.com/essandess/isp-data-pollution}}: The goal of this method is to eliminate the mutual information between actual browsing history and the manipulated one. According to~\cite{ye2009noise} mutual information vanishes if:
	\begin{align}
		n\textunderscore {Noise} \geq (n\textunderscore {Calls} - 1) \times n\textunderscore{PossibleCall}
	\end{align}
	where $n\textunderscore {PossibleCall}$ is the number of domains that a user might visit per day, and $n\textunderscore {Calls}$ is the number of visited domains. For instance, if a user visits 100 domains and requests 200 calls per day, then \textit{ISPPolluter} adds 20,000 links randomly to the history. We choose this method to see if eliminating mutual information can preserve privacy in practice.
\end{itemize}
\begin{figure}[t]
	\centering
	\small
	{\includegraphics[scale=0.45]{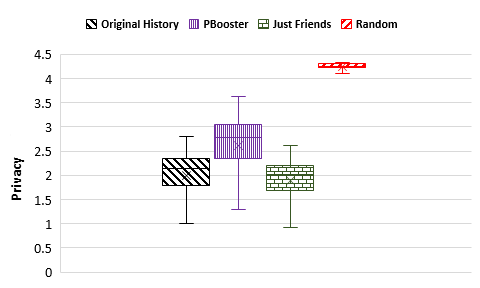}}
	\caption{\textbf{Privacy distributions before and after running anonymization techniques.}}\label{figprivacy-change}
\end{figure}
\subsection{Privacy Analysis}

To answer the first question, we first compare each user's privacy before and after anonymization for browsing histories with size 100 ($|\mathcal{H}^u|=100$). Fig.~\ref{figprivacy-change} depicts box plots of the distributions of users' privacy measured using Eq.\ref{eq:privacy}. The privacy-utility trade-off coefficient $\lambda$ is also fixed to $\lambda=10$. Results demonstrate how privacy increases after deploying \textsc{PBooster} in comparison to \textsc{JustFriends} approach and original history. This shows that adding links from friends cannot make significant change in privacy. This is because of Homophily effect~\cite{mcpherson2001birds}. The \textsc{Random} technique leads to the most uniform topics distribution and thus highest privacy among others. 

We now evaluate the efficiency of \textsc{PBooster} against the de-anonymization attack introduced in~\cite{su2017anonymizing}. We measure the attack success rate by the metric $\mathcal{X}\%=\frac{n_c}{N}\times100$ where $n_c$ is the total number of users that have been successfully mapped to their Twitter accounts, and $N$ indicates the total number of users in the dataset. We consider the attack as successful if the user is among the top 10 results returned by the attack. Lower values of this measure translates to the higher privacy and stronger defense. We evaluate all methods on histories with different sizes. The results for browsing histories with different $\lambda$ are demonstrated in Fig.~\ref{Fig:privacy}. Note due to the lack of space, we have removed the similar trend that we observed for $|\mathcal{H}^u|=30$. We observe the following:
\begin{figure}[t]
	\centering
	\small
	\subfloat[Browsing history of size 50]{\includegraphics[scale=0.39]{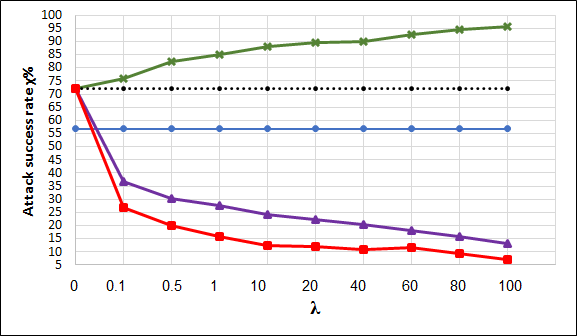}}\quad
	\subfloat[Browsing history of size 100]{\includegraphics[scale=0.39]{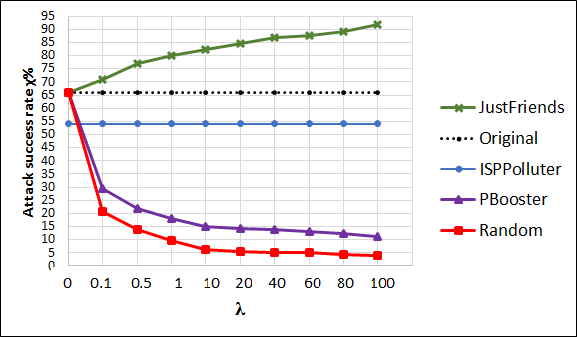}}
	\caption{\textbf{Attack Success rate for different sizes of history.}}\label{Fig:privacy}
\end{figure}

\begin{itemize}[leftmargin=*]
	\item \textsc{ISPPolluter} does not work properly in practice and is not robust to the attack which leverages traces of users' activities in social media. This confirms the idea of selecting links from non-friend users which inhibits the adversary to find the targeted user. 
	\item \textsc{Random} is more robust to the attack than \textsc{PBooster} and \textsc{JustFriends}. This demonstrates that adding random links from non-friends could perform better in terms of privacy.
	\item \textsc{JustFriends} decreases the privacy in comparison to the original history. This aligns well with the observations of~\cite{su2017anonymizing} suggesting that adding links from friends can even decrease the privacy.
	\item  Attack success rate decreases to $15\%$ after applying \textsc{PBooster}. We conclude that the generated history from \textsc{PBooster} is more robust to the attacks in comparison to original history and those generated from \textsc{JustFriends} and \textsc{ISPPOlluter}. This confirms the effectiveness of \textsc{PBooster} for preserving privacy.
	\item \textsc{PBooster} performs better when $|\mathcal{H}^u|=100$. This means larger history can help \textsc{PBooster} to model user's interests better and manipulate the history accordingly.
	\item \textsc{PBooster} is much more robust than \textsc{JustFriends}. This clearly shows the efficiency of the link selection approach.
	\item In \textsc{PBooster},the attack success rate first decreases with the increase of $\lambda$ and then it gets almost stable (for $\lambda\geq10$). This makes the selection of $\lambda$ easier and suggests that the privacy will not increase significantly after some point, confirming that adding more links does not always necessarily lead to more privacy.
	\item By deploying \textsc{PBooster}, the attack success rate decreases even when $\lambda$ slightly changes from 0 to 0.1, which confirms the effectiveness of \textsc{PBooster} in anonymizing browsing histories.
\end{itemize}
\begin{figure}[t]
	\centering
	\small
	\subfloat[Browsing history of size 50]{\includegraphics[scale=0.39]{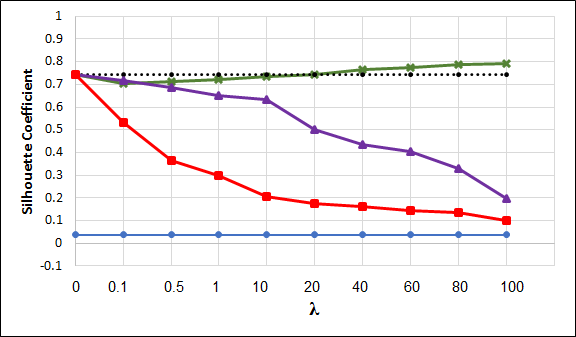}}\quad 
	\subfloat[Browsing history of size 100]{\includegraphics[scale=0.39]{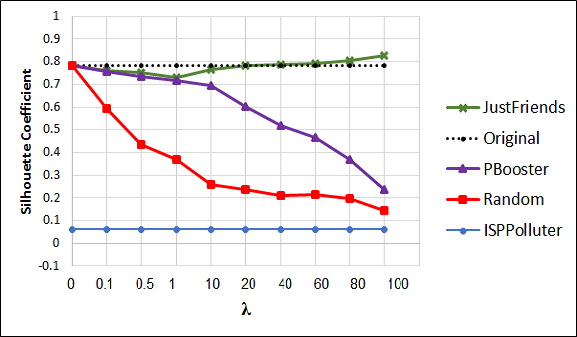}}
	\caption{\textbf{Silhouette coefficient after $k$-means with $k=5$ for different sizes of history.}}\label{Fig:utility}
\end{figure}
\begin{table}[h]
	\centering
	\small
	\caption{\textbf{Attack success rate after applying \textsc{PBooster} for different values of $h$ with $\lambda=10$.}}\label{Tab:privacy2}
	\begin{tabular}{|l|l|l|l|l|l|}\hline
		& $h=5$ & $h=15$ & $h=25$ & $h=50$ & $h=100$ \\ \hline 
		$\mathcal{X}$ & 27.83 & 19.58 & 15.13 &7.83 & 5.33 \\ \hline
	\end{tabular}
\end{table}

To study the effect of $h$ (size of batches of links in browsing history), we repeat the attack with different values of $h$ for $|\mathcal{H}^u|=100$ with $\lambda=10$ which was empirically found to work well in our problem. Results are demonstrated in Table~\ref{Tab:privacy2} suggesting that increasing $h$ can help to model users' preferences more accurately and \textsc{PBooster} can further decrease the traceability of users by making the profiles more ambiguous. Although this increases the privacy, it increases the anonymization waiting time which could result in sudden publishing of history without proper anonymization. 

\subsection{Utility Analysis}

To answer the second question, we investigate the utility of the manipulated histories to estimate the change in quality of services. We evaluate the utility of manipulated history via a well-known machine learning task, i.e. clustering. Prior works~\cite{ungar1998clustering,sarwar2002recommender} have indicated the benefits of applying clustering in personalization which can help to offer similar services to same cluster of people.

We use $k$-means to cluster users into  $k=5$ groups based on their topic preferences distribution $\hat{p_u}$. We evaluate the utility of browsing histories according to the quality of generated clusters via Silhouette coefficient. Silhouette coefficient ranges from $[-1,1]$, where a higher value indicates better clusters while a negative value indicate that a sample has been assigned to the wrong cluster. Values near zero indicate overlapping clusters (i.e., all users are similar to each other). The results are demonstrated in Fig.\ref{Fig:utility}. The same trend was observed for $|\mathcal{H}^u|=30$ but we remove it due to space limitations. We make the following observations:

\begin{itemize}[leftmargin=*]
	\item Clusters by \textsc{ISPPolluter} has the lowest Silhouette coefficient close to $0$ (i.e., clusters are almost overlapping). This shows that adding a large number of random links results in making all users similar to each other and thus severe utility degradation.
	\item The quality of clusters formed by \textsc{Random} decreases by increasing $\lambda$. This confirms that adding links randomly decreases the utility of browsing history and thus shows the importance of the topic and link selection phases.
	\item \textsc{JustFriends} can even increase the utility of the manipulated browsing history. This is not surprising and the reason is that friends have more similar tastes to each other than random people (Homophily effect~\cite{mcpherson2001birds}). Therefore, adding links from a friends' history will not change the preferences distributions significantly. Utility also improves slightly with increase in value of $\lambda $.
	\item Generated history by \textsc{PBooster} has better quality when $|\mathcal{H}^u|=100$ in comparison to $|\mathcal{H}^u|=50$. This shows that \textsc{PBooster} works better when more user's information is fed to it.
	\item The quality of clusters by \textsc{PBooster} decreases with increase in value of $\lambda$. The change is even sensible when $\lambda\geq20$. 
	\item The quality of data generated by \textsc{PBooster} is comparable to the original data when $\lambda\leq 10$. Moreover, \textsc{PBooster} reaches the optimal point in privacy-utility trade-off by fixing $\lambda=10$. 
\end{itemize}
\begin{table}[h]
	\centering
	\small
	\caption{\textbf{Silhouette coefficient after applying \textsc{PBooster} for different values of $h$ with $\lambda=10$.}}\label{Tab:utilit2}
	\begin{tabular}{|l|l|l|l|l|l|}\hline
		& $h=5$ & $h=15$ & $h=25$ & $h=50$ & $h=100$ \\ \hline 
		$S$ &0.477 & 0.5699 & 0.694 & 0.731 & 0.762 \\ \hline
	\end{tabular}
\end{table}

We repeat $k$-means with different values of $h$ for $|\mathcal{H}^u|=100$ with $\lambda=10$. Results are demonstrated in Table~\ref{Tab:utilit2} and suggest that increasing $h$ will lead to more accurate representation of users and thus improvement in the utility of data. However, as discussed earlier, the main drawback with increasing value of $h$ is increasing the risk of sudden history publishing without proper anonymization. 

\subsection{Privacy-Utility Trade-off}
\begin{figure}[t]
	\centering
	{\includegraphics[scale=0.45]{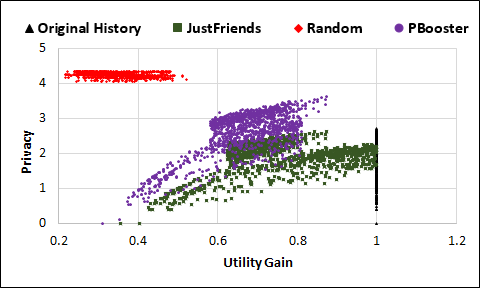}}
	\caption{\textbf{Privacy vs utility gain for different approaches.}}\label{Fig:tradeoff}
\end{figure}
To answer the third question, we plot the privacy and utility gain values for each user after applying different approaches over histories with size 100. We measure the privacy by Eq.\ref{eq:privacy} and utility gain as $1-utility\textunderscore {loss}$ using the Eq.\ref{eq:uitlity}. Different colors and markers represent different approaches. Each marker represents a user, with measures over his manipulated history with $h=25$ and $\lambda=10$.
\begin{itemize}[leftmargin=*]
	\item The original history gains the utility of 1 and the privacy to some extent. \textsc{Random} reaches the highest privacy but loses utility. \textsc{JustFriends} results in higher data utility gain in comparison to other methods but reaches a lower level of privacy. The result of \textsc{PBooster} varies for different users, achieving different levels of privacy and utility according to their original browsing behavior, whereas all users gain similar level of privacy by \textsc{Random}.
	\item Users achieve higher privacy with \textsc{PBooster} than the original data comparing with other approaches. The achieved utility by \textsc{PBooster} is more than the utility by \textsc{Random} but less than the utility by \textsc{JustFriends}. The reason lies at the intrinsic trade-off between utility and privacy--higher privacy results in less utility. 
\end{itemize}

We compare the privacy and utility of browsing history manipulated by different techniques demonstrated in Fig.\ref{Fig:privacy} and Fig.\ref{Fig:utility}:
\begin{itemize}[leftmargin=*]
	\item \textsc{JustFriends} achieves the highest utility among all approaches while it is the most vulnerable method. \textsc{Random} approach is the most robust technique against de-anonymization attack, however has the most utility lost. \textsc{PBooster} provides high privacy but can sacrifice utility for high values of $\lambda$ ($\lambda \geq20$). 
	\item \textsc{PBooster} is the most efficient approach in terms of both privacy and utility. Setting $\lambda=10$, it returns the highest possible privacy while maintaining comparable utility with the original data.
\end{itemize}

\section{Conclusion and Future Work}
The need arises for users to protect their sensitive information such as browsing history from potential adversaries. 
Some users resort to Tor, VPN and HTTPS to remove their traces from browsing history to assure their privacy. However, these solutions may hinder personalized online services by degrading the utility of browsing history. In this study, we first quantified the trade-off between user privacy and utility and then proposed an efficient framework \textsc{PBooster} to address the problem of anonymizing web browsing histories while retaining the utility. Our experiments demonstrate the efficiency of the proposed model by increasing the user privacy and preserving utility of browsing history for future applications. 
In future, we would like to investigate personalized utility-privacy trade-off, by tweaking framework parameters to fit specific needs of each user. We also plan to replicate the work by exploring other mechanisms for anonymizing web browsing histories.
Last but not least, we would also like to collect real-world data and investigate the efficiency of \textsc{PBooster} in terms of both privacy and utility in practice.
\begin{acks}
	 This material is based upon the work supported, in part, by NSF \#1614576, ARO W911NF-15-1-0328 and ONR N00014-17-1-2605.
\end{acks}

\end{document}